\documentclass[twocolumn,prl,showpacs,superscriptaddress]{revtex4}
\usepackage{graphicx}
\usepackage{dcolumn}
\usepackage{bm}

\begin{document}

\title{Demonstration of Geometric Landau-Zener Interferometry in a
Superconducting Qubit}
\author{Xinsheng Tan}
\affiliation{National Laboratory of Solid State Microstructures,
School of Physics, Nanjing University, Nanjing 210093, China}
\affiliation{Department of Physics and Astronomy, University of
Kansas, Lawrence, KS 66045, USA}

\author{Dan-Wei Zhang}
\affiliation{Laboratory of Quantum Engineering and Quantum
Materials, SPTE, South China Normal University, Guangzhou 510006,
China}

\author{Zhentao Zhang}
\thanks{Present address: School of Physics Science and Information
Engineering, Liaocheng University, Liaocheng 252059, China.}
\affiliation{National Laboratory of Solid State Microstructures,
School of Physics, Nanjing University, Nanjing 210093, China}

\author{Yang Yu}
\email{yuyang@nju.edu.cn}
\affiliation{National Laboratory of Solid State Microstructures, School of Physics,
Nanjing University, Nanjing 210093, China}

\author{Siyuan Han}
\email{Han@ku.edu}
\affiliation{Department of Physics and Astronomy, University of Kansas, Lawrence, KS
66045, USA}

\author{Shi-Liang Zhu}
\email{slzhu@scnu.edu.cn}
\affiliation{National Laboratory of Solid State Microstructures, School of Physics,
Nanjing University, Nanjing 210093, China}
\affiliation{Laboratory of Quantum Engineering and Quantum Materials, SPTE, South China
Normal University, Guangzhou 510006, China}

\begin{abstract}
Geometric quantum manipulation and Landau-Zener interferometry
have been separately explored in many quantum systems. In this
paper, we combine these two approaches to study the dynamics of a
superconducting phase qubit. We
experimentally demonstrate Landau-Zener interferometry based on
the pure geometric phases
 in this solid-state qubit.
We observe the interference caused by a pure geometric phase
accumulated in the evolution between two consecutive Landau-Zener
transitions, while the dynamical phase is cancelled out by a
spin-echo pulse. The full controllability of the qubit state as a
function of intrinsically robust geometric phase provides a
promising approach for quantum state manipulation.
\end{abstract}

\maketitle

Solid-state devices based on superconducting Josephson junctions
present an excellent platform for exploring quantum mechanics and
hold promise for applications in quantum information processing
\cite{Makhlin}. Great successes have been achieved in
demonstrating the quantum coherent nature of
these artificial atoms \cite%
{Nakamura,Vion,Yu,Chiorescu,Claudon,McDermott,Berkley,Steffen,Plantenberg,Wallraff,Chiorescu04,Mayer,Sillanpaa,Wilson,Paik,Rigetti,Vijay}%
. However, in order to realize practical quantum computation, much
research is still needed to find the optimal balance between
controllability and decoherence.

A promising approach to achieve robust control of a quantum system
is by using geometric phases  which arise from a cyclic evolution
of the quantum system
\cite{Berry,Zanardi,Sjoqvist,Ekert,Jones,Duan,Leibfried,Abdumalikov,Falci}.
These geometric phases have been observed in a wide variety of
systems including superconducting qubits
\cite{Leek,Mottonen,Abdumalikov}.
In particular, the robustness of geometric phases against certain noises \cite%
{Chiara,Zhu2005} has been confirmed in a recent experiment based
on a superconducting qubit \cite{Berger}. One of the most
promising applications of geometric phases on quantum control is
geometric quantum computation. The design of fault-tolerant
quantum logic gates is the central issue in realizing quantum
computation. The phases in a quantum gate are usually a
combination of geometric phases and dynamical phases. The
performances of these two types of phases are essentially
different: in contrast to dynamical phases which are sensitive to
the rate of passage, geometric phases depend solely on the global
geometry of the trajectory traversed by the state of the system.
Because geometric phases are immune to certain local fluctuations,
it was proposed that quantum gates with purely geometric phases,
which are called geometric quantum gates \cite{Zanardi,Ekert}, are
intrinsically fault-tolerant \cite{Jones,Duan,Falci,Sjoqvist}.
Interestingly, high-fidelity geometric quantum gates have been
experimentally realized by several groups
\cite{Leibfried,Abdumalikov}.

On the other hand, Landau-Zener (LZ) transition (LZT) has been
demonstrated to be another important approach to control a quantum
system \cite{Nori,Damski,Oliver,Sillanpaa2,Petta, Sun}. When a
quantum two-level system is driven through an avoided energy-level
crossing, the system can
undergo nonadiabatic transitions with the LZT probability%
\begin{equation}
P_{LZ}=\exp (-\pi \Delta ^{2}/2\hbar v),  \label{Plz}
\end{equation}%
where $\Delta $ is the energy gap at the avoided crossing and $v$
is the speed of energy variation \cite{Nori}. Quantum interference
and manipulations of qubit states based
on this phenomenon have been reported in different quantum systems \cite%
{Nori,Damski,Oliver, Sillanpaa2,Petta, Sun}. However, in these
studies, the dynamical phases dominate the entire evolution. In
other words, while LZ interferometry and geometric quantum
manipulation are both well explored in many quantum systems, they
have not been jointly examined in detail. In a recent interesting
paper \cite{GLZ}, the manipulation of geometric phases with LZT in
a superconducting charge qubit was theoretically proposed;
however, no experimental observation has been reported so far.

In this Letter, we fill this gap by reporting an experiment that
combines the LZT and geometric phase manipulation. We design and
experimentally realize LZ interferometry with pure geometric
phases in a superconducting phase qubit. Since the dynamics of a
superconducting qubit coupled to microwave fields is analogous to
the dynamics of a spin-1/2 particle subjected to a rotating
magnetic field, all kinds of geometric phases may accumulate in a
designed evolution \cite{Berry}. We observe the oscillations of
the excited state population versus the geometric phase, which is
an ubiquitous signature of quantum interferences. Our numerical
simulation results using the measured energy relaxation and phase
decoherence times agree well with the experimental data,
confirming the observation of {\it the geometric LZ interference}.
Furthermore, the observed contrast of LZ interferences with pure
geometric phases is higher than that of dynamical LZ
interferences, which shows the robustness of the geometric LZ
interferometry (GLZI). Since LZ
interferences have been proposed to have many applications in various quantum systems \cite%
{Nori}, the demonstrated GLZI in our experiment paves a new way to
achieve reliable control of qubit states. Therefore, our work
should shed light on and stimulate interests in applying LZ
interferometry for quantum state control in general and
fault-tolerant quantum information processing in particular.

The system used to demonstrate GLZI is a flux-biased
superconducting phase qubit.   The truncated Hamiltonian of the lowest two levels ($\left\vert 0\right\rangle $, $%
\left\vert 1\right\rangle $)  in
the energy bases is $H_{q}=\frac{1}{2}\hbar \omega \sigma _{z}$ \cite{Martinis02,Sun}, where $%
\sigma _{z}$ is the Pauli operator in the $z$ direction  \cite%
{Supplementary}. $\hbar \omega $ denotes the energy difference
between $\left\vert 0\right\rangle $ and $\left\vert
1\right\rangle $, which could be tuned by varying the external
magnetic flux $\phi _{ex}$ threading the loop, as illustrated in
Fig. 1(a). Moreover, the state of qubit can be controlled by
microwave
(MW) irradiation. With a MW field, the Hamiltonian becomes $%
H^{\prime }=\frac{1}{2}\hbar \omega \sigma _{z}+\hbar \Omega \cos (\omega
_{m}t+\theta )\sigma _{x},$ 
where $\Omega $ is the Rabi frequency proportional to the
amplitude of the MW, $\omega _{m}$ ($\theta$) is the frequency
(phase) of the MW. With respect to the frame
rotating with frequency $\omega _{m}$, we apply the rotating wave approximation to obtain 
\begin{equation}
H_{\mathcal{B}}=\frac{1}{2}\hbar \delta \sigma
_{z}+\frac{1}{2}\hbar \Omega _{x}\sigma
_{x}+\frac{1}{2}\hbar \Omega _{y}\sigma _{y}\equiv \hbar \mathcal{\vec{B}\cdot }%
\frac{\mathbf{\vec{\sigma} }}{2},  \label{HB}
\end{equation}%
where $\delta =\omega -\omega _{m}$ is the detuning, $\Omega
_{x}=\Omega \cos \theta $, and $\Omega _{y}=\Omega \sin \theta $.
The dynamics of the qubit is thus identical to that of a spin-1/2
particle subjected to an effective magnetic field
$\mathcal{\vec{B}=(}\Omega _{x},\Omega _{y},\delta )$. Therefore,
the geometric phases may accumulate in a designed evolution
\cite{Berry}.

\begin{figure}[tbph]
\vspace{0.3cm}\ \includegraphics[width=8.2cm]{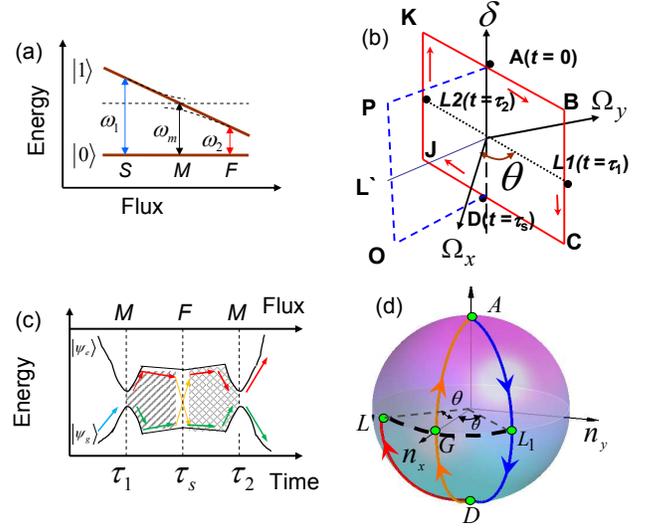} 
\caption{(Color online)  (a) The qubit
eigenenergies (solid lines) $E_{0,1}$ (with respect to ground state energy $E_{0}$%
) versus flux bias. An effective avoided crossing (dashed lines)
related to $H_{\mathcal{B}}$ is created at point $M$ by the MW
with frequency $\protect\omega _{m}$. (b) Effective magnetic
fields. The trajectory along ABCDJKA (ABCDOPA) is considered in
this letter (Ref. \cite{GLZ}). (c) The corresponding instantaneous
eigenvalues of $H_{\mathcal{B}}$ versus time and flux bias. The
qubit is initially in the ground state $|\protect\psi _{g}\rangle
=|0\rangle $, which splits at time $t=\protect\tau _{1}$ during
the first LZT, evolving along two different paths and interfering
at time $t=\protect\tau _{2}$ of the
second LZT. At point $F$, the state is flipped by a $\protect\pi $%
-pulse to eliminate the dynamical phase. (d) The Bloch vectors of
the qubit in the GLZI considered in Ref. \cite{GLZ} and here. }
\end{figure}

The Hamiltonian $H_\mathcal{B}$ is essentially identical to that
in Ref. \cite{GLZ} and can therefore be used to realize GLZI. In
the following, we briefly summarize GLZI proposed in Ref.
\cite{GLZ} and our modifications for implementation in a
superconducting phase qubit. In the original proposal in Ref.
\cite{GLZ}, the dynamical phase was not removed. Our method
improved upon this by eliminating both the dynamical phase and the
Stokes phase based on a spin-echo method, which was used to remove
dynamical phases in geometric quantum computation
\cite{Ekert,Jones}. To illustrate our scheme, we plot the
effective magnetic fields in Fig. 1(b), where the trajectory along
ABCDOPA is considered in Ref. \cite{GLZ} and the path along
ABCDJKA is studied here. In Fig. 1(c), we plot the energies of the
adiabatic states $|g\rangle$, $|e\rangle$ as a function of time in
a GLZI, obtained by instantaneous diagonalization of
$H_\mathcal{B}$. The LZTs occur at the avoided level crossings at
time $t=\tau_1,\ \tau_2$. The probability of a non-adiabatic LZT
at such a crossing is given approximately by Eq. (\ref{Plz}),
where $\Delta =\Omega $ and $v=\hbar (\omega _{1}-\omega
_{2})/\tau _{p}$, with $\tau _{p}$ being the LZ sweeping time
defined here as the time swept from point $S$ to point $F$ in Fig.
1(a). Here $\omega_1$ $(\omega_2)$ is the energy level spacing at
point $S$ $(F)$. Comparing our Fig. 1(b.c) with Fig. 1 (b,c) in
Ref. \cite{GLZ}, we make two modifications. One is that the
trajectory of effective field is now a rectangle. The other
modification, which is a substantial improvement, is the use of a
spin-echo  at time $\tau_s$ to remove the dynamical phase
\cite{Dynamical} as well as to make the evolution between two LZs
a closed path. To illustrate the latter point, we plot the
evolution path $n_j^g=\langle \psi_g|\sigma_j|\psi_g\rangle$
$(j=x,y,z)$ with initial ground state $|\psi_g(t=\tau_1)$ between
two LZTs on the Bloch sphere in Fig. 1(d) under the condition of
adiabatic approximation. The path $L_1  D L^\prime$ is for the
trajectory studied in Ref. \cite{GLZ},  where the geometric phase
is $\theta$ determined by the spherical triangle $L_1DL^\prime$.
On the other hand, the trajectory of $n_j^g$ considered here is
along the closed path $L_1DGAL_1$ in Fig. 1(d). The trajectory
$L_1D$ of the evolution follows the magnetic field $L_1 CD$ in
Fig. 1(b). The spin-echo leads the Bloch vector to evolve alone
$DGA$, and finally the Bloch vector evolves along $AL_1$ (as an
excited state) when the magnetic field moves along $DJL_2$ in Fig.
1(b). Therefore, the geometric phase accumulated between the two
LZ transitions is also $\theta$ determined by the trajectory $L_1
DGAL_1$ in Fig. 1(d).

Following the adiabatic-impulse method outlined in Ref.
\cite{GLZ,Nori,Damski}, we obtain the population on the excited
state after the entire evolution,
\begin{equation}
P_{1}=1-4P_{LZ}(1-P_{LZ})\sin ^{2}\theta.  \label{P1}
\end{equation}%
It is clear from Eq. (\ref{P1}) that $P_{1}$ is independent of the
dynamical phase and hence insensitive to fluctuations in qubit
transition frequency $\omega (t)=E_1(t)-E_0(t)$. Furthermore,
comparing with that in Ref. \cite{GLZ}, the intereference pattern
resulting from our method is independent of the Stokes phase
\cite{Kayanuma}. Therefore, our GLZI scheme should be more robust
against various harmful dynamical effects.

We use a trilayer Al/AlOx/Al superconducting phase qubit
fabricated on crystalline Al$_{2}$O$_{3}$ substrate to implement
our scheme experimentally. Figure 2(a) shows the principle
circuitry of the qubit and its measurement. The qubit is cooled to
30 mK in a dilution refrigerator. We measure the energy relaxation
time $T_{1}$ and decoherence time $T_{2}$ of the qubit using the
standard pump-probe and spin-echo, respectively \cite{Martinis02},
and the results are $T_{1}=118\pm 3$ ns and $T_{2}=157\pm 11$ ns
\cite{Supplementary}.

\begin{figure}[tbph]
\vspace{0.5cm} \includegraphics[width=8.2cm]{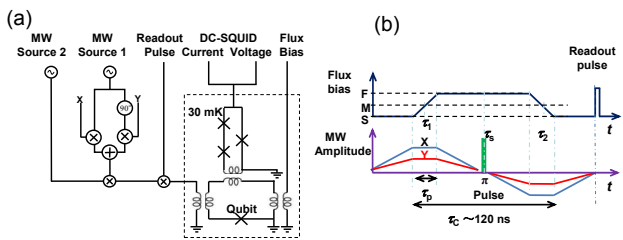} 
\caption{(Color online) (a) Schematic circuit of the phase qubit
and its experimental setup. The crosses in the dashed box
represent Josephson junctions. Magnetic flux bias line, microwave
and readout control lines, and DC-SQUID are inductively coupled to
the qubit. An IQ mixer is used to control the X and Y components
of MW1. (b) Schematics of the temporal profile of waveforms used
to perform GLZI. An avoided crossing is created at point $M$ by
MW1 (lower panel: blue and red lines). MW2 produces the
$\protect\pi $-pulse at $t=\protect\tau _{s}$ to null out the
dynamical phase. The symbols $S$, $M$, and $F$ correspond to those
in Fig. 1(a). }
\end{figure}

Figure 2(b) shows the temporal profile of the flux bias and MW
fields utilized to realize GLZI. In our experiment, the energy
level spacing at points $S$ and $F$ are $14.4$ GHz and $14.2$ GHz,
respectively [Fig. 1(a)]. The frequency of MW1, which creates the
avoided crossing, is $14.3$ GHz. The size of the energy gap
$\Omega $ at the avoided crossing can be tuned by adjusting the
amplitude of MW1. We control the Hamiltonian $H_{\mathcal{B}}$
along a closed loop in the field $\mathcal{\vec{B}}$ [Fig. 2(b)].
The $z$ component of the field is controlled by the flux bias, and
$\Omega _{x}$ and $\Omega _{y}$ are controlled using an IQ-mixer
by adjusting the voltage level applied to the I and Q channels of
the mixer. At $t=\tau _{s}$, we use MW2 to apply a $\pi $ pulse to
the qubit, which causes the qubit to evolve along the path $DGA$
in Fig. 1(d). The pulse width is only several nanoseconds, which
is much smaller than $\tau _{p}$, so it is reasonable to
approximate it by an instant $\pi $ pulse in our numerical
simulation. The $\theta$ angle in Fig. 1(b) and Fig. 1(d) is
determined by the ratio between the in-phase and quadrature
components of the microwave. Finally, after the second LZT is
completed, a short readout pulse is applied to make a projective
measurement of the qubit state.

\begin{figure}[tbph]
\includegraphics[width=3.8cm]{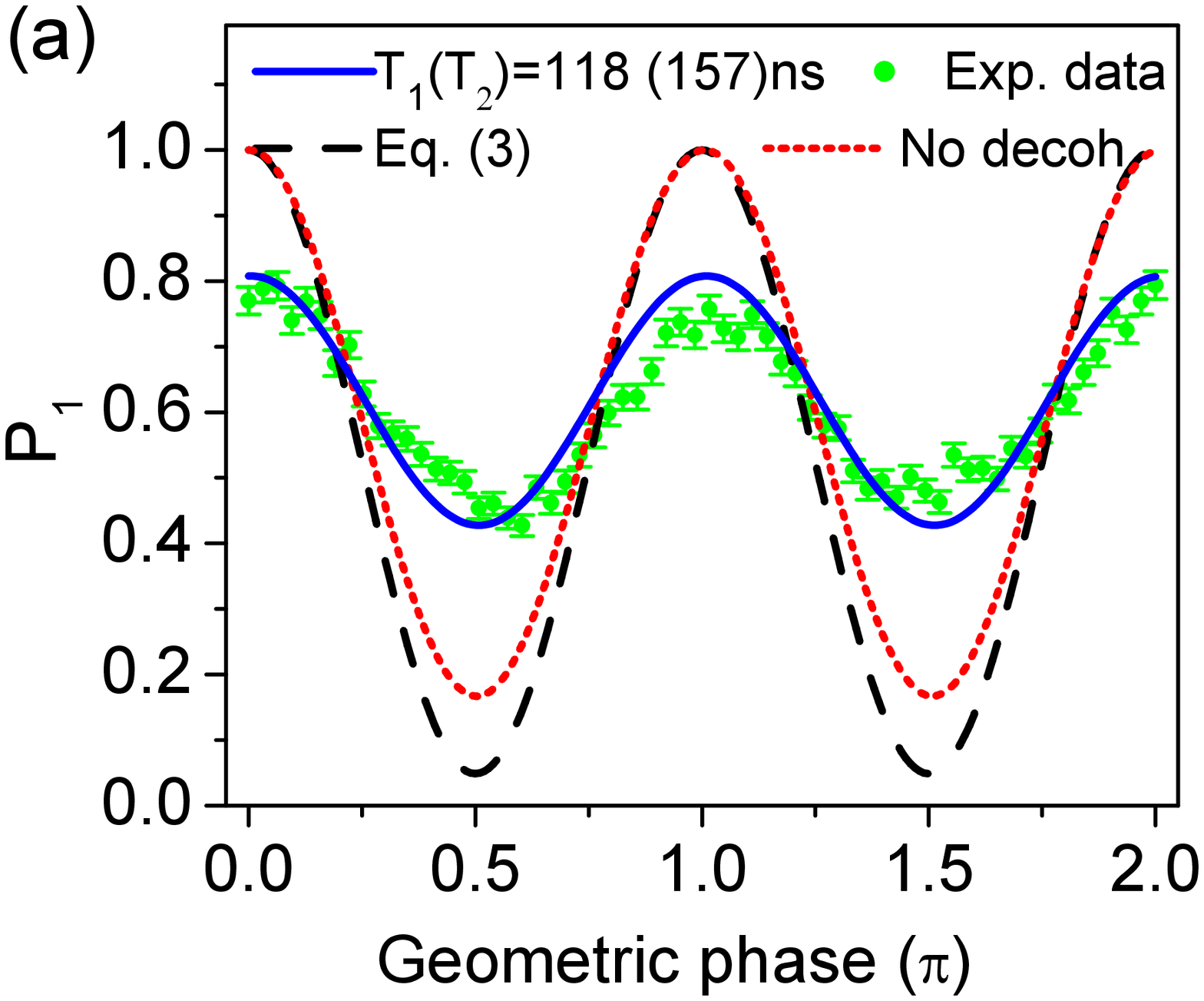}\ \ \ \ \ \ %
\includegraphics[width=3.8cm]{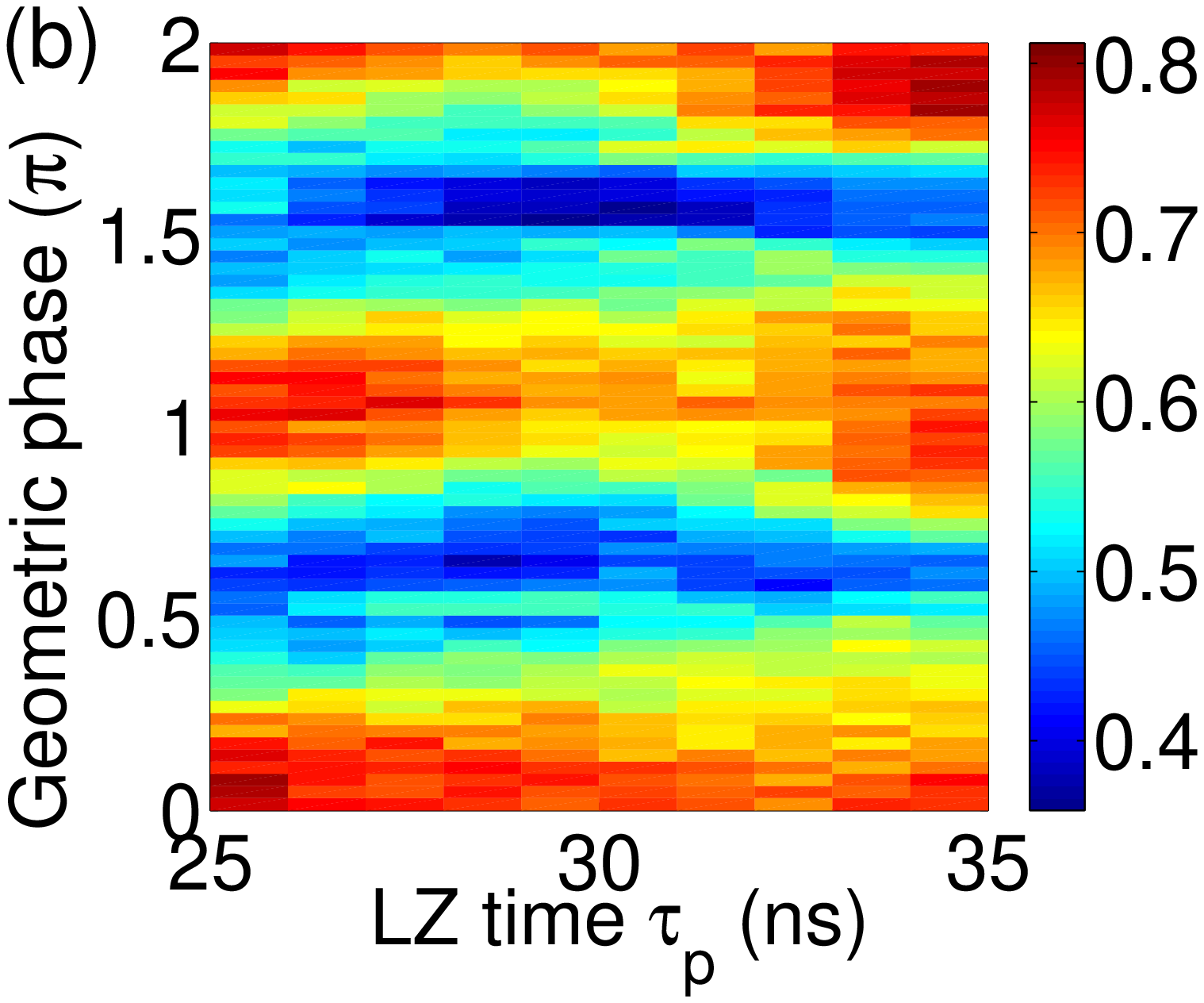}
\includegraphics[width=4cm]{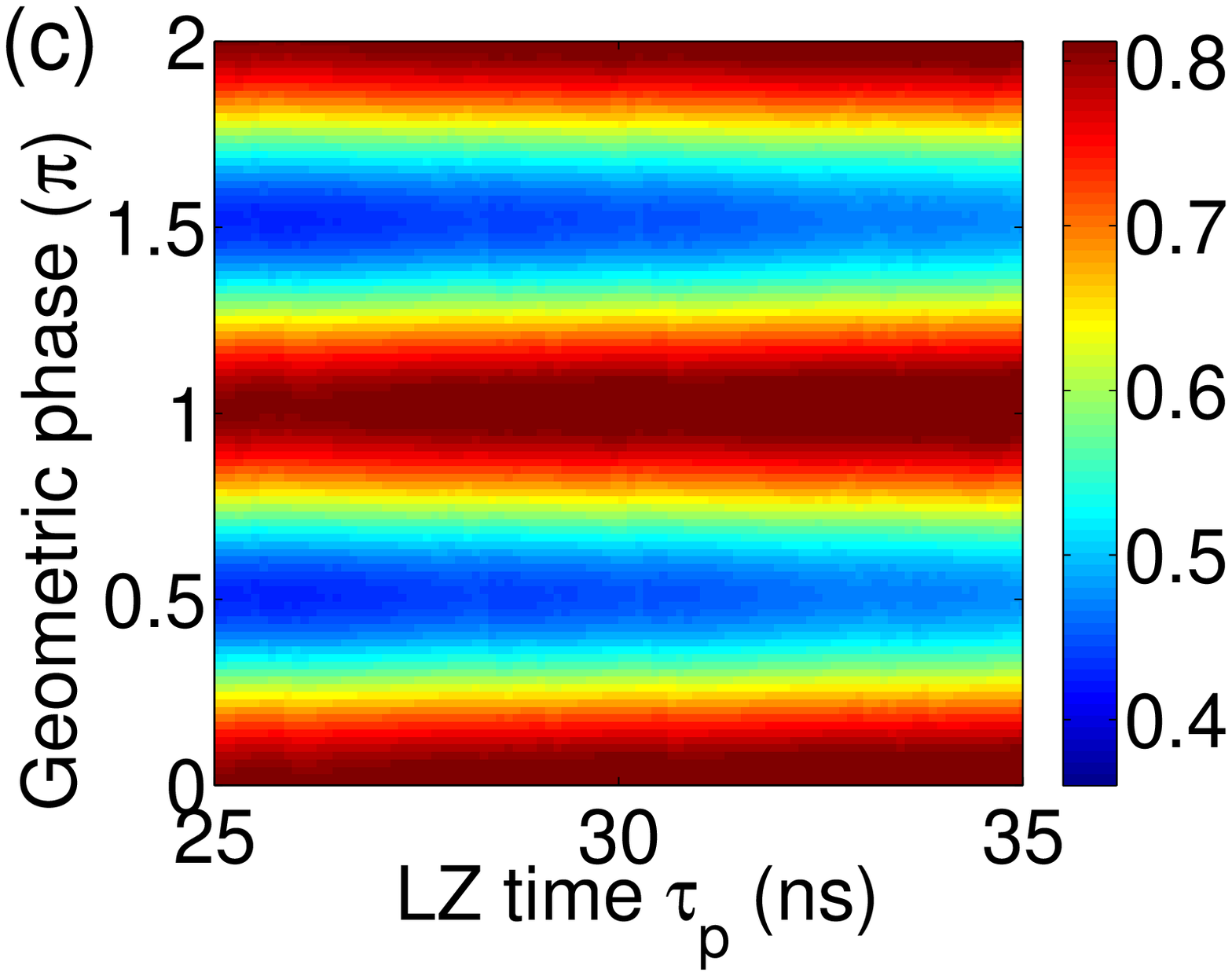}%
\includegraphics[width=4cm]{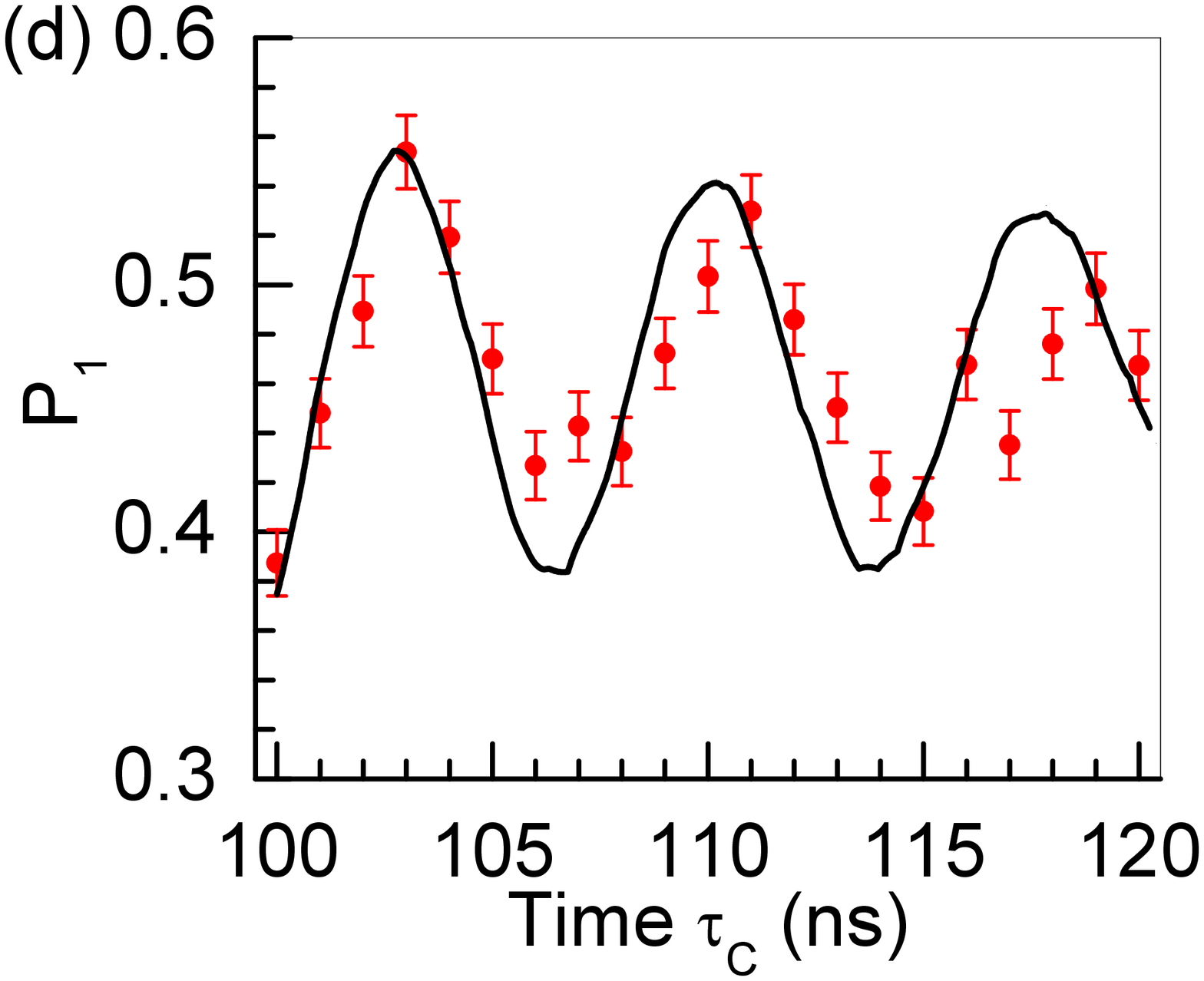}
\caption{(Color online) (a) Population $P_1$ versus geometric
phase with $\protect\tau_p=25$ ns and $\tau_C=100$ ns. Green dots
are experimental data,
while the black dashed line, the red dotted line, and the blue
solid line are theoretical results obtained from Eq. (\ref{P1}),
by numerically solving the Schr\"{o}dinger equation and the master
equation (\ref{MasterEq}) with decoherence $T_1=118 $ ns and
$T_2=157$ ns,
respectively. The color scale represents (b) measured $P_1$, and (c) simulated $P_1$ with the master equation versus geometric phase and LZ period time $%
\protect\tau_p$ for $\tau_C=100$ ns.  (d) Measured population
$P_{1}$ versus time $\protect\tau _{C}$ for $\protect\tau _{p}=25$
ns in dynamical LZ interferometry. The solid line is the numerical
simulation of the master equation.
Other parameters: $\Delta/2\protect\pi=\Omega/2\protect\pi= 20$
MHz, $\protect\delta/2\protect\pi=100$ MHz.}
\end{figure}

Figure 3 shows population $P_{1}$ as a function of the geometric phase and the LZ sweeping time $\tau _{p}$. As expected from Eq. (\ref%
{P1}), population $P_{1}$ is observed to oscillate sinusoidally as
a function of the geometric phase. However, as shown in Fig. 3(a),
the amplitude of the
observed oscillation is substantially smaller than that predicted by Eq. (\ref{P1}%
). We noticed that some approximations, such as the
adiabatic-impulse approximation and the asymptotic value of
$P_{LZ}$ given by Eq. (\ref{Plz}), are applied to obtain Eq. (3),
whereas the decoherence and energy relaxation are not considered.
To confirm that the observed oscillation is indeed caused by the
geometric phase of the system, in the following, we first discuss
in which regime our  experiment is performed. Since
$H_{\mathcal{B}}$ is cyclic with a period around $\tau _{C}\sim
100$ ns, which is defined as the time spent traversing the cyclic
path $SMFMS$, we may consider that the qubit is actually driven by
an ac field with frequency $\omega _{C}/2\pi =1/\tau _{C}\sim 10$
MHz. Thus our experiment is in the region of $\Delta /\omega
_{C}\sim 2$, whereas the previous experiments that
investigated LZTs are in the limits of $\omega _{C}>>\Delta $ \cite{Oliver} or $%
\omega _{C}<<\Delta $ \cite{Sillanpaa2}. In addition, to verify
the validity of those approximations used in Eq. (\ref{P1}),
we numerically solve the Schr\"{o}dinger equation with the Hamiltonian (\ref{HB}%
), and the result is shown as the red dotted line in Fig. 3(a).
Comparing this line with the black dashed line obtained from the
analytical expression given by Eq. (\ref{P1}) with $P_{\text
{LZ}}\simeq0.61$ (which is derived by Eq. (\ref{Plz}) for the
parameters $\Delta/2\pi=20$ MHz, $(\omega_1-\omega_2)/2\pi=200$
MHz and $\tau_p=25$ ns), the difference around $\theta =n\pi $
with $n$ an integer is negligible. On the other hand, around
$\theta =(n+1/2)\pi $ the difference between the numerical and
analytical results is relatively larger. To understand and resolve
this apparent discrepancy,  we plot the time evolution of
population $P_1$ in Fig. S4 in Supplementary Material by
numerically solving Schr\"{o}dinger equation for the cases of
$\theta=0$ and $\theta=\pi/2$, respectively. Population $P_1$
saturates after the first LZT, which is a clear-cut signature that
the adiabatic-impulse model is valid under the experimental
conditions. However, the saturation value $P_{LZ}^\prime
\simeq0.71$ is not equal to $P_{LZ} \simeq0.61$ as directly
derived from Eq.(\ref{Plz}). After we use $P_{LZ}^\prime$ to
replace  $P_{LZ}$ in Eq. (\ref{P1}), we obtain a line which is
almost identical to the red dotted line  in Fig. 3(a). Therefore,
we confirm that for our experiment,  the adiabatic-impulse model
is valid; however, the $P_{LZ}$ in Eq. (\ref{Plz}) should be
modified under the experimental conditions.

Now the question is why the oscillation amplitude of the
experimental result is significantly smaller than that of the
numerical (analytical) results while they have the same
oscillatory dependence of $P_{1}$ on the geometric phase $\theta$.
As demonstrated below, the quantitative discrepancy between the
experiment and calculations is due to energy relaxation and
decoherence. Because $T_{1}$ and $T_{2}$ of the qubit are
comparable to $\tau _{C}$, the effect of qubit decoherence cannot
be ignored. In order to have a meaningful comparison between the
experiment and theory, we take the effects of finite $T_{1}$ and
$T_{2}$ into consideration by numerically solving the master
equation. The quantum dynamics of the system is thus described by
the master equation of the time evolution of the density matrix
$\rho $:
\begin{equation}
\dot{\rho}=-\frac{i}{\hbar }[H_{\mathcal{B}},\rho ]-\Gamma \lbrack
\rho ], \label{MasterEq}
\end{equation}%
where the system Hamiltonian $H_{\mathcal{B}}$ is given by Eq.
(\ref{HB}), and the second term, $\Gamma \lbrack \rho ]$,
describes the relaxation and dephasing processes
phenomenologically. In a concrete expression, Eq. (\ref{MasterEq})
can be rewritten as $\dot{\rho}_{11}=-i[\frac{\Omega
}{2}(e^{-i\theta }\rho _{01}-e^{i\theta }\rho _{10})]-\Gamma
_{1}\rho _{11},$ $ \dot{\rho}_{10}=-i[\frac{\delta }{2}\rho
_{10}+\frac{\Omega }{2}e^{-i\theta }(\rho _{00}-\rho
_{11})]-\gamma \rho _{10},$ $\dot{\rho}_{00}=-\dot{\rho}_{11}$
with ${\rho }_{01}={\rho }_{10}^{\star }$, $\Gamma _{1}\equiv
1/T_{1}$, and $%
\gamma \equiv 1/T_{2}$.

The result obtained by solving the master equation without
decoherence
is identical to that obtained by numerically solving the Schr\"{o%
}dinger equation. Furthermore, the simulated result using the
measured qubit  $T_{1}=118$ ns and $T_{2}=157 $ ns,  shown as the
blue solid line in Fig. 3(a), agrees  fairly well with the
experimental data. In addition, from Eq. (\ref{Plz}), we know that
by adjusting the sweeping speed of the flux bias $v=\hbar (\omega
_{1}-\omega _{2})/\tau _{p}$, one could change the transition
probability $P_{LZ}.$ However, since the geometric phase does not
depend on $P_{LZ}$, the period and phase of $P_{1}(\theta )$
oscillation should  be insensitive to changes in $v$ [Fig. 3(c)].
In Fig. 3(b), we show the measured $P_{1}(\theta )$ oscillation
for the different LZI sweeping time $\tau _{p},$ which agrees
quite well with the numerical results. Moreover, unlike the LZ
interference of accumulated dynamical phases, $P_{1}$ does not
oscillate with $\tau _{p}$, indicating that the dynamical phase is
well eliminated by the spin-echo pulse.

Finally, we compare the performance of the geometric LZ
interference with that of the conventional dynamical LZ
interference. The measured  $P_{1}$ as a function of the cyclic
time $\tau _{C}$ for $\tau _{p}=25$ ns is shown in Fig. 3(d). For
dynamical LZ interferometry, the temporal profiles of the flux
bias and MW fields are the same as those depicted in Fig. 2(c),
except that no $\pi $ pulse is applied at point $F$. The phase
angle $\theta $ of the microwave is kept constant, while $\tau
_{C} $ is varied by changing the time spent in the plateau part of
Fig. 2(c). Contrary to that of the geometric LZ interference,
$P_{1}$ oscillates notably when
$\tau _{C}$ is varied. More importantly, for the similar range of  time $%
\tau _{C}\sim 100$ ns, the contrast (defined as the difference
between the maximum and the minimum amplitudes of the oscillation)
of the dynamical LZ interference is about 0.16, which is much
weaker than the corresponding value of 0.4 for the geometric LZ
interference. Because  $4P_{LZ}(1-P_{LZ})$ \cite{Nori} is also the
theoretical prediction for the contrast of dynamical LZ
interferences, this result demonstrates that the geometric phase
is more robust and stable against a variety of error-causing
dynamical processes such as those caused by microscopic two-level
systems coupled to the qubit \cite{Sun,Simmonds}.

In summary, we have experimentally demonstrated GLZI in a
superconducting qubit. Our results indicate that GLZI may have
advantages in many applications involving quantum state
manipulation \cite{Nori} such as simulating the time-reversal
symmetry with LZ interferometry \cite{Gustavsson}. Since our
geometric LZ approach combines the simplicity of LZ interferometry
and the robustness of geometric phases for quantum state
manipulation, it may very well open a new venue for fault-tolerant
quantum information processing.

\acknowledgments
 We thank X. Wen, G. Sun, B. Mao, and M. Gong
for their helpful discussion. This work is supported by the SKPBR
of China (2011CB922104, 2011CBA00200), NSFC (91021003,
11274156,11125417), NSF of Jiangsu (BK2010012), PAPD, and the
PCSIRT. S. Han is supported in part by NSF of United States
(PHY-1314861). We gratefully acknowledge Northrop Grumman ES in
Baltimore, MD, for foundry support and thank R. Lewis, A.
Pesetski, E. Folk, and J. Talvacchio for technical assistance. X.
T., D.-W. Z., and Z. Z. contributed equally to this work.

\newpage

\begin{appendix}

\section{Supplemental Material}

\section{S1: Truncated Hamiltanian of the flux-biased phase qubit}

The Hamiltonian of a flux-biased superconducting phase qubit with
the junction capacitance $C$ and the loop inductance $L$ reads
\begin{equation}
H=E_{C}\hat{q}^{2}-E_{J}\cos \hat{\phi}+E_{L}(\hat{\phi}-\phi
_{ex})^{2},
\end{equation}%
where the charging energy $E_{C}=\frac{(2e)^{2}}{2C}$, the Josephson energy $%
E_{J}=\frac{I_{c}\phi _{0}}{2\pi },$ and the inductance energy $E_{L}=\frac{1%
}{2L}(\frac{\phi _{0}}{2\pi })^{2},$ with $\phi _{0}$ being the
single flux quantum and $I_{c}$ denoting the critical current of
the junction. Here $q$ is the charge on the junction capacitor,
$\phi $ is the superconducting phase difference across the
junction, and $\phi _{ex}$ is the externally applied magnetic flux
threading through the superconducting loop of the phase qubit.
Quantum mechanically, the charge operator $\hat{q}$ and phase
difference $\hat{\phi}$ satisfy the commutator
$[\hat{q},\hat{\phi}]=1$ and form a set of degrees of freedom that
fully characterize the system. The potential energy part of the
above Hamiltonian is
\begin{equation}
U=-E_{J}\cos \hat{\phi}+E_{L}(\hat{\phi}-\phi _{ex})^{2}.
\end{equation}%
Given appropriate $E_{J}$ and $E_{L},$ when $\phi _{ex}=\pi $, it
is a symmetric double-well potential, with the states in two
potential wells corresponding to different circulation
super-currents. In order to operate in a phase qubit mode, the
external flux should be biased far away from the symmetric point
$\pi $. In this case, there exists a shallow potential well which
contains a few of quantized energy levels.

In order to estimate the energy level spacing of the phase qubit,
we first consider the classical dynamics of the above Hamiltonian.
It is equivalent to a particle of mass $C/4e^{2}$ moving in a
double well potential

\begin{equation}
U=-E_{J}\cos \phi +E_{L}(\phi -\phi _{ex})^{2}.
\end{equation}

The maximum and minimum points of the potential well can be
determined by the condition

\begin{equation}
\frac{\partial U}{\partial \phi }=E_{J}\sin \phi +2E_{L}(\phi
-\phi _{ex})=0.
\end{equation}%
The maximum is located at $\phi _{\max }=\pi $ for the symmetric situation $%
\phi _{ex}=\pi .$ For the minimum, $E_{J}\sin \phi _{\min
}=-2E_{L}(\phi _{\min }-\phi _{ex}),$ which can not be solved
analytically. However,
the small oscillation frequency near the minimum point is determined by $%
\omega=\sqrt{2E_C\frac{\partial ^{2}U}{\partial \phi ^{2}}|_{\phi
=\phi
_{min}}}$, i.e.,%
\begin{eqnarray}
\omega&=\sqrt{2E_C(E_{J}\cos \phi _{\min }+2E_{L})} &   \nonumber \\
&=\sqrt{2E_C(E_{J}\sqrt{1-\sin ^{2}\phi _{\min }}+2E_{L})} &   \nonumber \\
&=\sqrt{2E_C(E_{J}\sqrt{1-[2E_{L}(\phi _{\min }-\phi _{ex})/E_{J}]^{2}}%
+2E_{L})}. &
\end{eqnarray}

\setcounter{figure}{0}
\renewcommand{\thefigure}{S\arabic{figure}}
\begin{figure}[tbph]
\includegraphics[width=8cm]{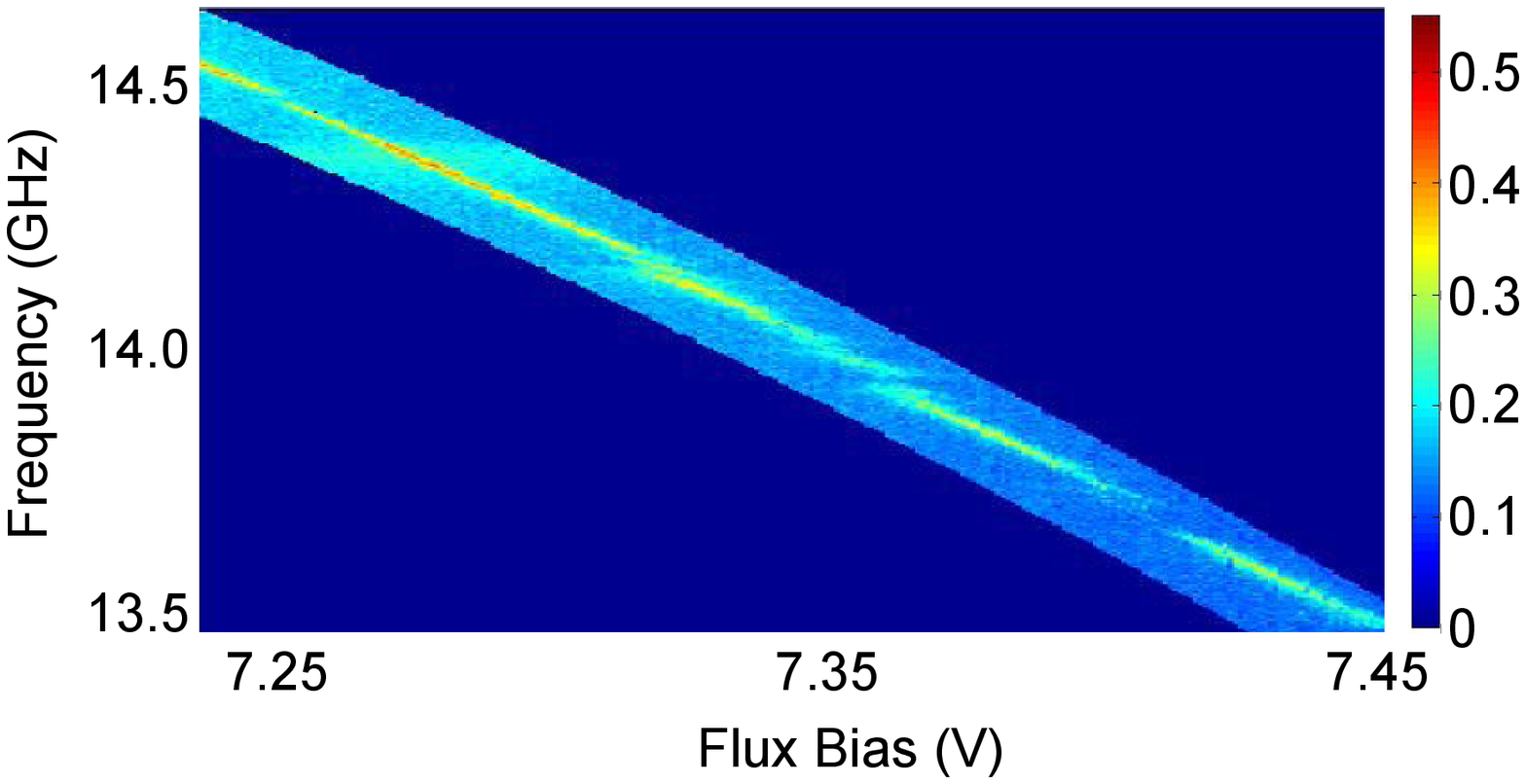}
\caption{(color online). Spectroscopy of the phase qubit used in
our experiments. By applying a voltage waveform with an arbitrary
wavefunction we can bias the qubit at working area (so the unit of
the horizontal axis is voltage) and initialize it to the ground
state. A long microwave pulse is used to generate transition to
the excited state. Then we measure the resonant peak as a function
of the flux bias. Some splittings caused by the coupling to the
microscopic two-level systems can be observed [1].}
\end{figure}


After quantization, the small oscillation frequency defines the
energy level spacing. Since $\phi _{\min }$ changes with $\phi
_{ex},$ the potential well
is anharmonic, causing the energy level spacing to be slightly different ($%
\omega _{21}\approx 0.9\omega _{10}$) [2]. Therefore, the two
lowest levels (denoted as 0, 1) can be isolated from the higher
levels and function as qubit. The truncated Hamiltonian of the
phase qubit in the energy bases now reads
\begin{equation}
H_{q}=\frac{1}{2}\hbar \omega \sigma _{z},
\end{equation}%
where $\sigma _{z}$ is the Pauli operator in orientation $z$. From
the expression (9) of $\omega $, it is clear that the energy level
of the qubit is experimentally tunable through the external
magnetic flux $\phi _{ex}$, which is observed in the spectroscopy
(Fig. S1).

\section{S2: The relaxation and decoherence times}

The energy relaxation time $T_1$ is independently measured using
the standard pump-probe method [2]. The qubit is initialized in
the ground state. At time $t = 0$, a $\pi$ pulse excites the qubit
to the excited state. After a time duration $t_{delay}$, a readout
pulse is applied to the qubit and the population on the excited
state $P_1$ is obtained (inset of Fig. S2). By fitting the
exponential decay of $P_1$ vs. $t_{delay}$, we obtain $T_1 = 118
\pm 3$ ns (Fig. S2). We also measured the dephasing time using the
spin-echo technique [3]. The pulse sequence is shown in the inset
of Fig. S3. The first $\pi/2$ pulse puts the qubit in the
equatorial plane. The system freely evolves for a short time
$\tau$. Before the completely randomization of the qubit phase, a
 $\pi$ pulse is applied to flip
the qubit state. The system inversely evolves the same amount of
time $\tau$, resulting in the refocus of the system. The second
$\pi/2$ pulse projects the system to the z axis. Then we apply a
readout pulse to measure the qubit state. As expected, the
population on the excited state decays with the $\tau$ or the
total time between two $\pi/2$ pulses. By fitting $P_1$  vs.
$t_{total}$, we obtain $T_2 = 157\pm 11$ ns (Fig. S3).

\begin{figure}[tbph]
\vspace{1.0cm}
\includegraphics[height=4cm]{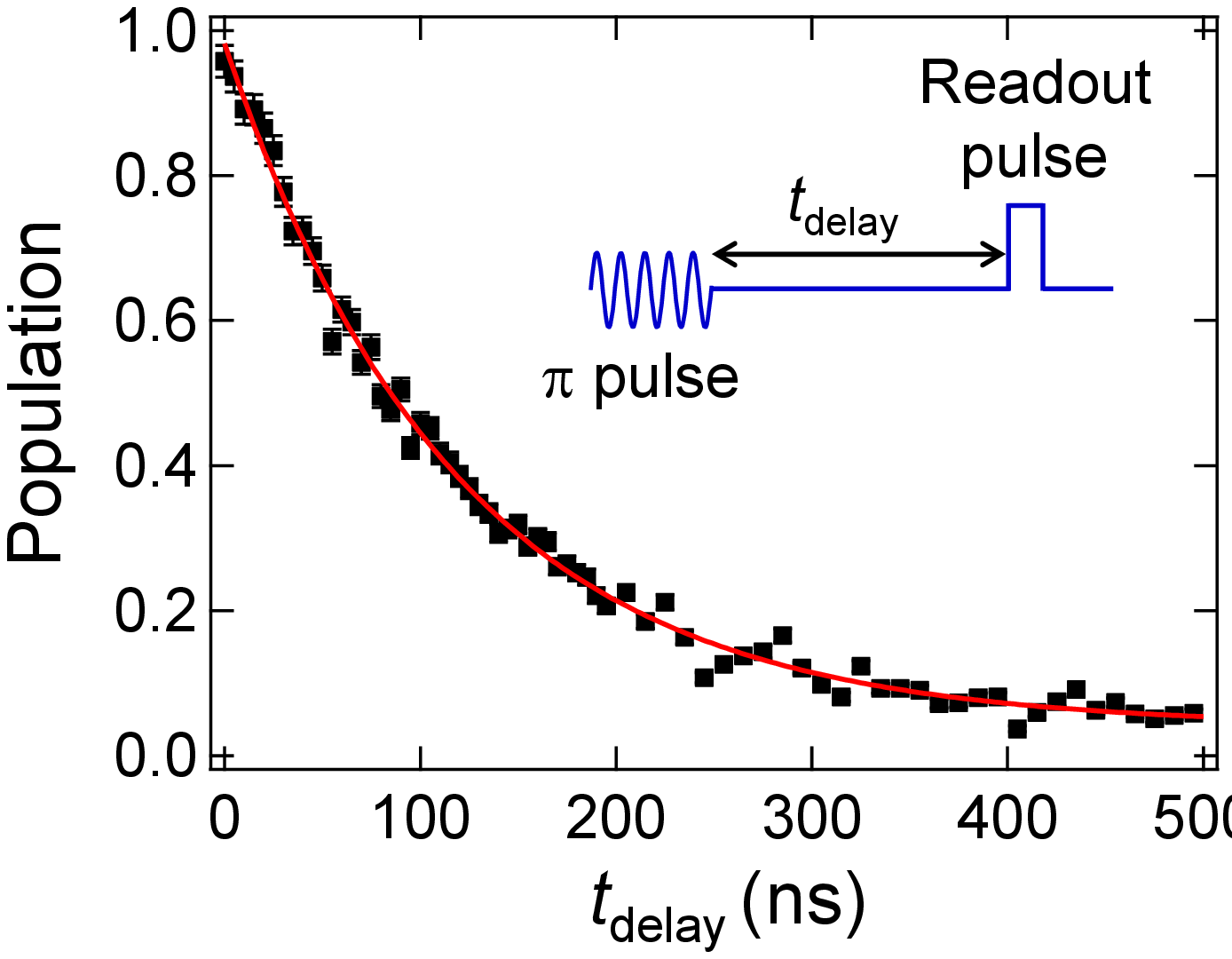}
\caption{(color online).  The population of the excited state
decay with time after the pumping microwave pulse is turned off.
Inset: schematic pulse sequence to measure $T_1$.}
\end{figure}
\begin{figure}[tbph]
\vspace{1.0cm}
\includegraphics[height=4cm]{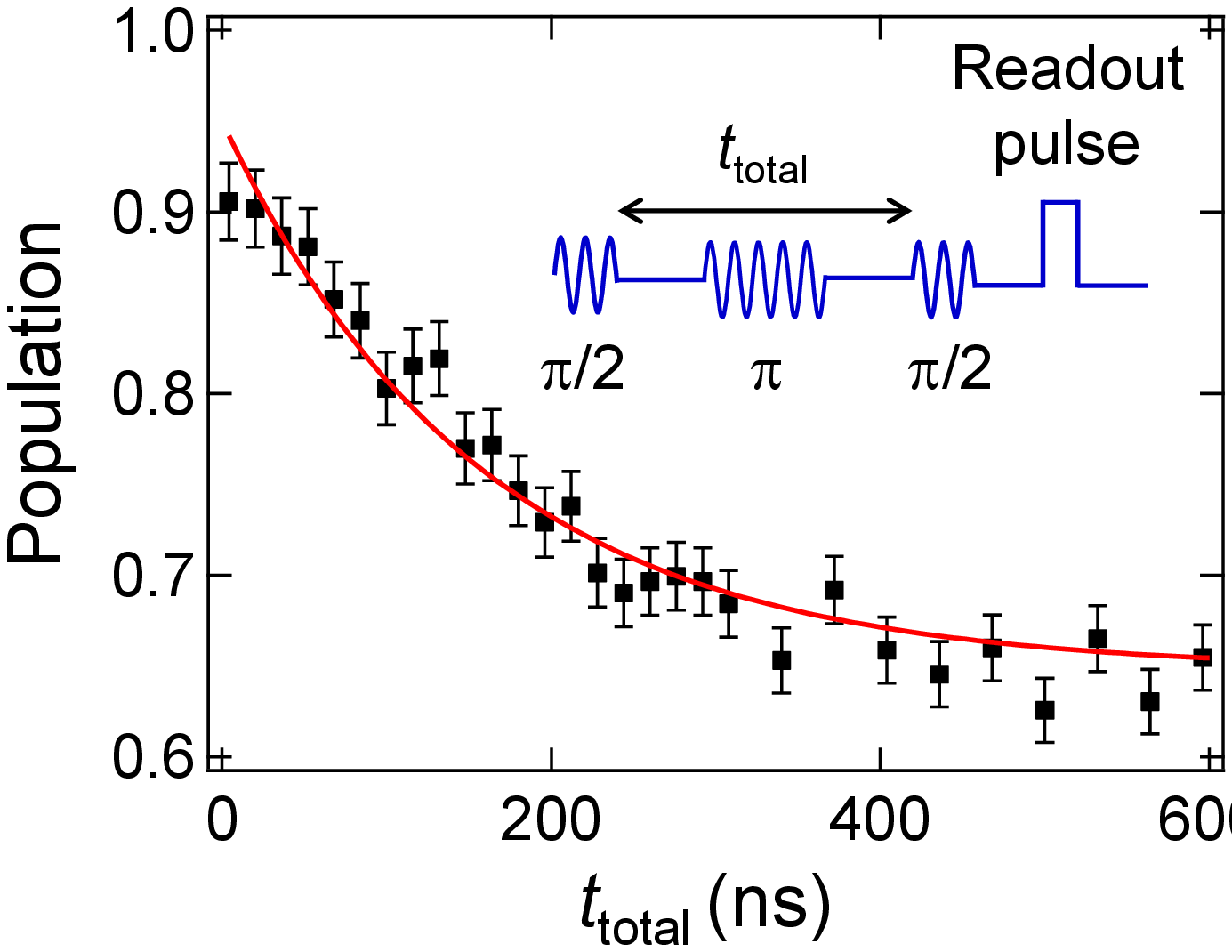}
\caption{(color online). The population of the excited state vs.
refocus time (time between two $\pi/2$ pulses). Inset: schematic
pulse sequence to measure $T_2$. }
\end{figure}

\section{S3: Time evolution of population $P_1$}


\begin{figure}[tbph]
\includegraphics[width=0.8\linewidth]{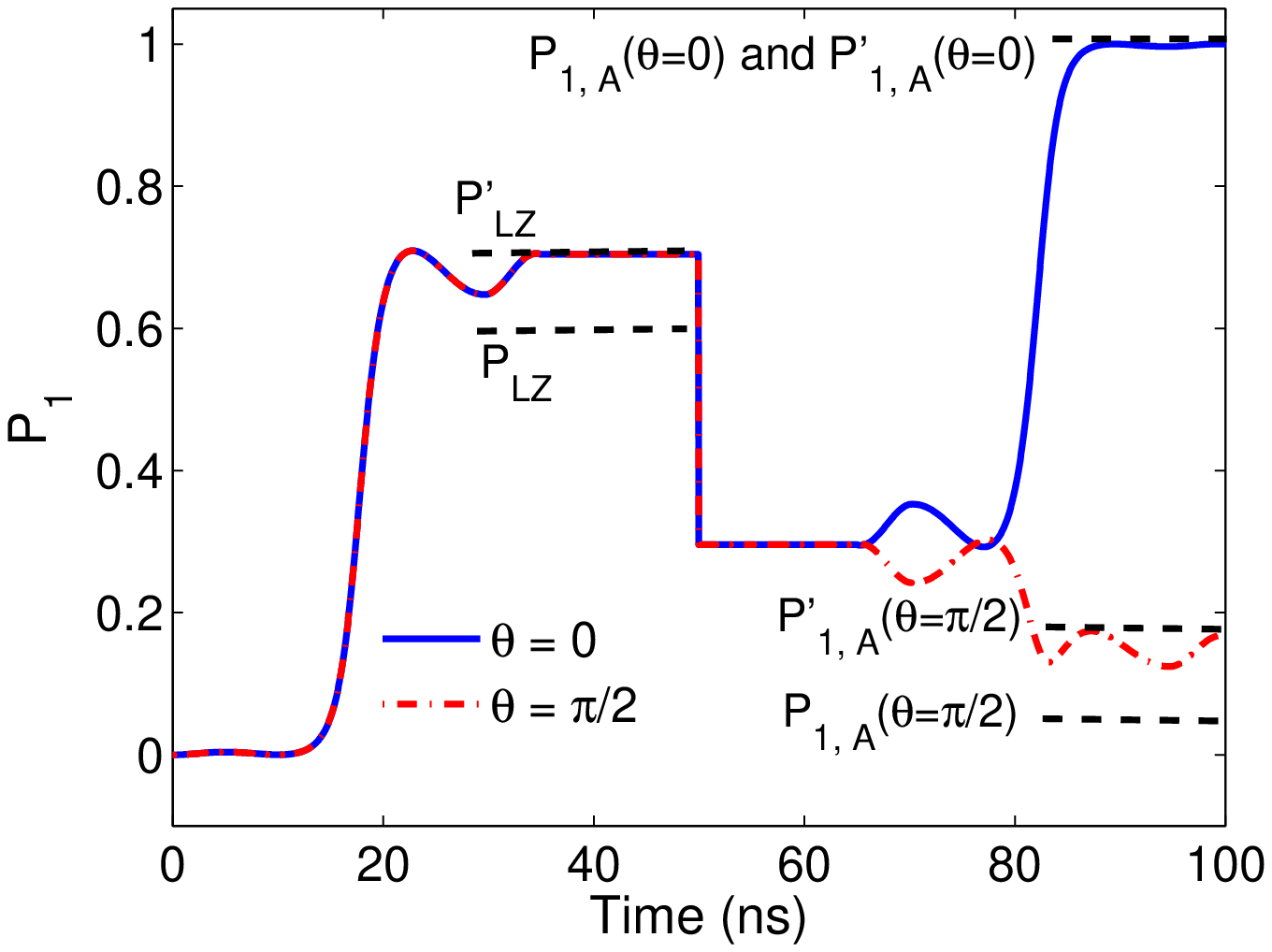}
\caption{(color online). Population of excited state $P_1$ as a
function of the evolution time by numerically solving the
Schr\"{o}dinger equation. The blue solid line and red dash-dotted
line are the numerical results for $\theta=0$ and $\theta=\pi/2$,
respectively. The inserted black dash lines to the left represent
$P_{LZ}\simeq0.61$ given by Eq. (1) and $P^\prime_{LZ}\simeq0.71$
the saturation value after the first LZ transition. The other
three black dash lines to the right represent the analytical
results of $P_1$ at the end of the evolution, which are given by
Eq. (3) with $P_{\text {LZ}}$ and $P'_{\text {LZ}}$. The
parameters are the same as those
in Fig. 3 in the main text.
}
\end{figure}

In Fig. 3(a) in the main text, the difference between the
numerical result by solving the Schr\"{o}dinger equation and the
analytical result by using Eq. (3) (with $P_{LZ}$ given by Eq.
(1)) is negligible at $\theta=n\pi$ (with $n$ an integer). But the
difference is slightly larger at $\theta=(n+1/2)\pi$. The
analytical expression Eq. (3) is actually based on two
approximations: the adiabatic-impulse approximation and the
approximation of the LZ transition probability Eq. (1). In this
section, we study the time evolution of population $P_1$ by
numerically solving the time-dependent Schr\"{o}dinger equation.
It is found that the adiabatic-impulse model is valid, but instead
of using Eq. (1), the $P_{LZ}$ value in Eq. (3) should be
determined from numerical calculation under the experimental
conditions.

To illustrate this argument, we plot in Fig. S4 the time evolution
of population $P_1$ for the cases of $\theta=0$ and
$\theta=\pi/2$, respectively, as well as the analytical result of
$P_1$ for comparison. Population $P_1$ saturates after the first
LZ transition, it is a clear indicator that the adiabatic-impulse
model is valid under the experimental conditions. However, the
saturation value $P_{LZ}^\prime \simeq0.71$, which is not equal to
$P_{LZ} \simeq0.61$ as directly calculated from Eq.(1) for the
parameters $\Delta/2\pi=20$ MHz, $(\omega_1-\omega_2)/2\pi=200$
MHz and $\tau_p=25$ ns. This deviation confirms that Eq. (1) is an
asymptotic formula which is only valid when both the starting and
ending points are far away from the avoided crossing [4]. We have
numerically checked that the deviation is decreased by setting
parameters, such as increasing $\delta$ by two times, one may have
$P_{\text {LZ}} \approx P'_{\text {LZ}}$.

For $\theta=0$, the analytical result of $P_1$ at the end of the
evolution, as described by Eq. (3), is always one despite the LZ
transition probability. This agrees with the numerical result. For
$\theta=\pi/2$, the numerical result disagrees with the analytical
one if we use the LZ transition probability $P_{\text
{LZ}}\simeq0.61$ given by Eq. (1). However, the numerical and
analytical results give nearly equal values of $P_1$ if we use the
LZ transition probability $P^\prime_{\text {LZ}}\simeq0.71$
established by the numerical result (i.e. the saturation value
$P_{LZ}^\prime$ after the first LZ transition in Fig. S4).

\vspace{1cm}

\emph{References:} \vspace{0.2cm}

[1] R. W. Simmonds \textit{et. al.}, Phys. Rev. Lett. \textbf{93},
077003 (2004).

[2] J. Martinis, S. Nam, J. Aumentado, and C. Urbina, Phys. Rev.
Lett. \textbf{89}, 117901 (2002).

[3] E. L. Hahn, Phys. Rev. 80, 580 (1950).

[4] S. Shevchenko, S. Ashhab, and F. Nori, Phys. Rep. \textbf{492}%
, 1 (2010).

\end{appendix}


\begin{thebibliography}{99}
\bibitem{Dynamical}
The spin-echo pulse divides the process between two LZTs into two
symmetric segments. The dynamical phase accumulated in each
segment, which is proportional to the shaded area between points
$M$ and $F$ in Fig. 1(a), are equal in magnitude but have opposite
signs. Therefore, the dynamical phases between two LZs vanish.
\end{thebibliography}
\end{document}